\documentclass[prd,10pt,twocolumn,nofootinbib,preprint,superscriptaddress]{revtex4}
\pdfoutput=1
\usepackage[normalem]{ulem}
\usepackage{amstext}
\usepackage{amssymb}
\usepackage{amsmath}
\usepackage{graphicx}
\graphicspath{{plots/}}
\usepackage{url}
\usepackage{color}
\usepackage{caption}
\usepackage{subcaption}
\usepackage{placeins} 
\usepackage{ulem}
\usepackage[utf8]{inputenc}
\pdfoutput=1
\usepackage{textcomp}
\usepackage{comment}
\usepackage{yfonts}
\usepackage{epsfig,amsfonts,mathrsfs,amsmath,amssymb,graphicx,color,slashed,multirow}
\usepackage{amsmath,latexsym,amssymb,graphicx,slashed,color,enumerate,url,cancel,gensymb}
\usepackage{textcomp}

\usepackage[x11names]{xcolor}
\usepackage[colorlinks]{hyperref}

\usepackage{booktabs}
\usepackage{adjustbox}

\definecolor{vdrgreen}{rgb}{0.0, 0.7, 0.0}

\newcommand{\qtransfer}{\left|\mathbf{q}\right|}
\def\cevns{CE$\nu$NS }

\definecolor{indianred}{rgb}{0.8, 0.36, 0.36}
\definecolor{blue(ncs)}{rgb}{0.0, 0.53, 0.74}
\def\d{\mathrm{d}}
\AtBeginDocument{\hypersetup{citecolor=indianred,linkcolor=indianred,urlcolor=indianred}}

\usepackage{float}

\usepackage[T1]{fontenc}
\usepackage{ae,aecompl}
\graphicspath{{Figures/}}
\usepackage{appendix}



\newcommand{\AddrIISERB}{Department of Physics, Indian Institute of Science Education and Research - Bhopal, \\ 
Bhopal Bypass Road, Bhauri, Bhopal 462066, India}

\newcommand{\AddrAHEP}{%
  Institute of Experimental Physics, University of Hamburg, 22761, Hamburg, Germany}

\usepackage{orcidlink}
\begin{document}

\title{\LARGE \textcolor{indianred}{Can Elastic Neutrino Scattering Account for the LZ230616 Event?}}
\author{Ayan Chattaraj~\orcidlink{0009-0001-3561-7049}}
\email{ayan23@iiserb.ac.in}
\affiliation{\AddrIISERB}
\author{Anirban Majumdar~\orcidlink{0000-0002-1229-7951}}\email{anirban19@iiserb.ac.in}
\author{Dimitrios K. Papoulias~\orcidlink{0000-0003-0453-8492}}\email{dimitrios.papoulias@uni-hamburg.de}
\affiliation{\AddrAHEP}
\affiliation{\AddrIISERB}
\author{Rahul Srivastava~\orcidlink{0000-0001-7023-5727}}
\email{rahul@iiserb.ac.in}
\affiliation{\AddrIISERB}

\begin{abstract}
Recently, the LUX–ZEPLIN (LZ) Collaboration reported an isolated nuclear-recoil event at $248\pm23_{\rm stat}\pm23_{\rm syst}~\mathrm{keV_{nr}}$, in a region where the expected background is very small and conventional elastic dark matter (DM)–nucleus scattering cannot readily account for such a localized feature. We investigate whether LZ230616 could instead originate from coherent elastic neutrino–nucleus scattering (CE$\nu$NS). We consider neutrino–nucleus interactions within and beyond the Standard Model, together with exotic neutrino fluxes from DM annihilation ($\chi \chi \to \nu \bar{\nu}$) or decay ($\chi \to \nu \bar{\nu}$) into neutrino pairs and from primordial black hole (PBH) evaporation. We show that kinematic considerations, the accompanying low energy recoil spectrum, and existing constraints exclude a viable interpretation of LZ230616 in terms of elastic neutrino–nucleus scattering for all scenarios considered.
\end{abstract}

\maketitle
\section{Introduction}
A plethora of astrophysical and cosmological observations provides compelling evidence for the existence of dark matter (DM), which accounts for approximately 25\% of the energy density of the Universe~\cite{Planck:2018vyg}. Nevertheless, its particle nature and possible nongravitational interactions remain yet unknown. Nonrelativistic weakly interacting massive particles (WIMPs) are among the most motivated DM candidates~\cite{Lee:1977ua} 
and have been extensively searched for by numerous direct-detection experiments using different techniques. Leading searches employ dual phase liquid xenon time projection chambers
~\cite{XENON:2023cxc,PandaX:2024qfu,LZ:2024zvo}
and semiconductor detectors
~\cite{SuperCDMS:2018mne,EDELWEISS:2022ktt,CDEX:2023vvc,
SENSEI:2023zdf,DAMIC-M:2025luv},
while emerging low-threshold technologies, including superfluid helium and gaseous
detectors~\cite{SPICE:2023tru,QUEST-DMC:2023nug,NEWS-G:2023qwh},
aim to extend sensitivity to lighter DM candidates. These experiments have targeted rare nuclear- and electron-recoil signals at energies below approximately $100~\mathrm{keV}$, collectively probing a wide range of DM masses and interaction mechanisms. Despite these extensive efforts, no conclusive evidence for nongravitational DM interactions has yet been established.

Recently, the LUX–ZEPLIN (LZ) Collaboration~\cite{LZ:2026axp} extended its nuclear-recoil search to higher energies and, using an exposure of \(2.84~\mathrm{ton \times yr}\), reported a single unexplained candidate event at $E_{\rm nr}=248\pm23_{\rm stat}\pm23_{\rm syst}~\mathrm{keV}$. The LZ230616 event lies in a region where the expected background is very small and has a global significance of $2.6~\sigma$.
 Elastic processes such as coherent elastic neutrino–nucleus scattering (CE\(\nu\)NS) and conventional WIMP–nucleus scattering generally produce a recoil spectrum dominated by low-energy events and therefore does not naturally account for such an isolated high-energy feature. Inelastic processes, by contrast, can introduce a kinematic threshold and concentrate the signal within a restricted recoil-energy interval. Several interpretations based on endothermic or exothermic DM scattering have consequently been proposed~\cite{Fan:2026kxx,Visinelli:2026kgt,McCabe:2026crm,Smirnov:2026aqk,deLima:2026shq,Gu:2026vto,Dent:2026bji,Baer:2026fpy,Wang:2026ytg,Okada:2026eol}. Other proposed mechanisms involve fermionic DM absorption~\cite{Lou:2026idn},  boosted dark sector particles~\cite{Liang:2026coz,Alhazmi:2026efz,Kannike:2026qyl}, and  a possible neutrino-induced explanation based on atmospheric neutrino upscattering~\cite{Jeesun:2026vzo}.

In this work, we investigate whether LZ230616 could have a neutrino origin, considering both Standard Model (SM) and beyond the Standard Model (BSM) interactions. Simple kinematic considerations already severely restrict the neutrino sources capable of producing such a signal. In elastic neutrino–xenon scattering, generating a nuclear recoil within the $225 - 271~\mathrm{keV}$ interval requires a neutrino energy of at least $E_\nu\simeq118~\mathrm{MeV}$. Solar neutrinos are therefore kinematically unable to produce the event, while the diffuse supernova neutrino background (DSNB) is negligible at the required energies. Atmospheric neutrinos extend to sufficiently high energies and hence constitute the only conventional neutrino source with the necessary kinematic reach. Their flux, however, decreases rapidly with energy, while the large momentum transfer associated with the LZ230616 event strongly suppresses the coherent nuclear response. As a result, the expected SM event rate is far too small to account for the LZ observation. However, the presence of BSM light mediators would distort the event rate and potentially yield the observed recoil. We thus examine the possibility of whether novel scalar or vector interactions, i.e., the two scenarios with largest change in event rate, can sufficiently enhance the atmospheric neutrino contribution without overproducing lower energy recoils. Finally, we consider additional high energy neutrino fluxes originating from nonconventional neutrino sources such as from DM annihilation or decay and from primordial black hole (PBH) evaporation, and test whether they can reproduce the observed event while remaining compatible with the lower energy LZ data.

\section{CE$\nu$NS in the SM and beyond}
\cevns is a process in which a neutrino scatters coherently
off the entire nucleus. Within the SM, the differential cross section
with respect to the nuclear recoil energy $T_\mathcal{N}$
reads~\cite{Freedman:1973yd, Drukier:1984vhf}
\begin{equation}
\frac{\d\sigma^{\rm SM}_{\nu\mathcal{N}}}{\d T_\mathcal{N}}
= \frac{G_F^2 m_\mathcal{N}}{\pi}\,\left(Q_W^V\right)^2
\left[1 - \frac{m_\mathcal{N} T_\mathcal{N}}{2E_\nu^2}\right]\,,
\label{eq:cevns-sm}
\end{equation}
with $m_\mathcal{N}$ the nuclear mass, $E_\nu$ the incoming neutrino energy,
and the weak charge
\begin{equation}
Q_W^V = \mathbb{Z}\,\mathcal{F}_p\,g_p^V
+ \mathbb{N}\,\mathcal{F}_n\,g_n^V\,,
\end{equation}
where $g_p^V = 1/2 - 2\sin^2\theta_W$ and $g_n^V = -1/2$ are the proton and
neutron vector couplings, $\mathbb{Z}$ ($\mathbb{N}$) is the proton (neutron)
number, and $\qtransfer = \sqrt{2 m_\mathcal{N} T_\mathcal{N}}$ is the
three-momentum transfer. We fix $\sin^2\theta_W = 0.23857$~\cite{ParticleDataGroup:2022pth}  and incorporate  nuclear structure effects via shell model nuclear form factors  $\mathcal{F}_{p,n}(\qtransfer^2)$ for protons and neutrons~\cite{Fitzpatrick:2012ix}.

New physics can modify this cross section through a neutral mediator that couples to both the
neutrinos and quarks. For a vector mediator $Z'$ with mass $M_{Z'}$, the new
amplitude shares the similar Lorentz structure of the SM neutral current and hence
interferes with it
~\cite{Bertuzzo:2021opb, Majumdar:2024dms}
\begin{equation}
\frac{\d\sigma^{\rm SM+Z'}_{\nu\mathcal{N}}}{\d T_\mathcal{N}}
= \left[1 + \frac{Q_{Z'}\,\textsl{g}_V^2}
{\sqrt{2}\,G_F\,Q_W^V\left(M_{Z'}^2 + 2 m_\mathcal{N} T_\mathcal{N}\right)}
\right]^2
\frac{\d\sigma^{\rm SM}_{\nu\mathcal{N}}}{\d T_\mathcal{N}}\,,
\label{eq:cevns-vector}
\end{equation}
whereas a scalar mediator $\phi$ with mass $M_\phi$ does not interfere with
the SM and contributes additively~\cite{Farzan:2018gtr},
\begin{equation}
\frac{\d\sigma^{\phi}_{\nu\mathcal{N}}}{\d T_\mathcal{N}}
= \frac{m_\mathcal{N}^2\,T_\mathcal{N}\,Q_\phi^2}
{4\pi E_\nu^2 \left(M_\phi^2 + 2 m_\mathcal{N} T_\mathcal{N}\right)^2}\,.
\label{eq:cevns-scalar}
\end{equation}
The effective nuclear charges induced by the two mediators are
\begin{align}
Q_{Z'} &= \textsl{g}_{Z^\prime}^2Q^\nu_V\left[\mathbb{Z}\,\mathcal{F}_p\left(2Q^u_V + Q^d_V\right)
+ \mathbb{N}\,\mathcal{F}_n\left(Q^u_V + 2Q^d_V\right)\right]\,,
\\[1ex]
Q_{\phi} &= \textsl{g}_\phi^2\left[
\mathbb{Z}\,\mathcal{F}_p\sum_{q}\frac{m_p}{m_q}f^p_{T_q}
+ \mathbb{N}\,\mathcal{F}_n\sum_{q}\frac{m_n}{m_q}f^n_{T_q}
\right]\,,
\end{align}
where $f^{p,n}_{T_q}$ denote scalar nucleon form
factors~\cite{Freeman:2012ry, DelNobile:2021wmp} and $q=\{u,d\}$. For the
vector case we adopt a universal mediator that couples with equal strength to
neutrinos and to first generation quarks, $Q^\nu_V = Q^u_V = Q^d_V = 1$. We assume
universal couplings to the first generation quarks and quote our results in
terms of the effective couplings $\textsl{g}_{Z^{\prime},\phi} = \sqrt{g^\nu_{Z^{\prime},\phi}\,
g^{q}_{Z^{\prime},\phi}}$, since the cross sections depend on the neutrino  and
quark  couplings only through their product. Both
Eqs.~(\ref{eq:cevns-vector}) and (\ref{eq:cevns-scalar}) reduce to a
contact-like $\textsl{g}^2/M^2$ behavior once
$M_{Z^{\prime},\phi}^2 \gg 2 m_\mathcal{N} T_\mathcal{N}$, while for
$M_{Z^{\prime},\phi}^2 \ll 2 m_\mathcal{N} T_\mathcal{N}$ the recoil dependence of the
propagator becomes the dominant effect.

\section{Event rate and analysis strategy} 
For a given neutrino flux and interaction channel \(X\), the predicted recoil spectrum as a function of the reconstructed nuclear recoil energy \(T_\mathcal{N}^{\rm reco}\) is
\begin{eqnarray}
\frac{\d R}{\d T_\mathcal{N}^{\rm reco}}\bigg|_X
& = & {\cal E} N_T \,
\mathcal{A}(T_\mathcal{N}^{\rm reco})
\int \d T_\mathcal{N}\;
G\big(T_\mathcal{N}^{\rm reco}, T_\mathcal{N}\big) \nonumber \\
&& \times \int_{E_\nu^{\rm min}} \!\! \d E_\nu\;
\frac{\d\Phi_\nu}{\d E_\nu}\,
\frac{\d\sigma_X}{\d T_\mathcal{N}}\,.
\label{eq:rate}
\end{eqnarray}
Here, \(\mathcal E=2.84~\mathrm{ton\,yr}\) is the exposure, \(N_T\) is the number of target nuclei per ton of natural xenon, and \(\mathcal A(T_\mathcal{N}^{\rm reco})\) is the experimental efficiency. The minimum neutrino energy required to produce a recoil \(T_\mathcal{N}\) is denoted by \(E_\nu^{\rm min} = \sqrt{m_\mathcal{N} T_\mathcal{N}/2}\). The energy resolution is modeled by a Gaussian smearing function \(G(T_\mathcal{N}^{\rm reco},T_\mathcal{N})\) with width
$\sigma(T_\mathcal{N}) = 1.46
\sqrt{T_\mathcal{N}}\, 
\mathrm{keV_{nr}}$.
At the reconstructed energy of LZ230616, this parametrization reproduces the reported statistical uncertainty, \(\sigma(248~\mathrm{keV_{nr}})\simeq23~\mathrm{keV_{nr}}\). We construct the recoil energy binning by setting the width of each bin equal to twice the resolution evaluated at its center. The bin centered at \(248~\mathrm{keV_{nr}}\) therefore spans $225-271~\mathrm{keV_{nr}}$ and defines our signal region. We extend the same binning prescription to lower recoil energies, truncating the lowest bin at the analysis threshold \(T_\mathcal{N}^{\rm reco}=5.4~\mathrm{keV_{nr}}\).
 We identify the highest-energy bin, which contains the single candidate event, as the signal region, and we require the corresponding integrated
rate to reproduce it
\begin{equation}
R_{\rm I} = \int_{225~\mathrm{keV_{nr}}}^{271~\mathrm{keV_{nr}}}
\d T_\mathcal{N}^{\rm reco}\;
\frac{\d R}{\d T_\mathcal{N}^{\rm reco}}\bigg|_X = 1\,.
\label{eq:RI}
\end{equation}
Reproducing the event in the signal region is necessary but not sufficient:
any mechanism producing one event at $\sim 250~\mathrm{keV_{nr}}$ must
simultaneously remain consistent with the absence of any excess in the
low-energy region, where the experiment is far more sensitive. We therefore
also compute the rate integrated over the remaining bins,
\begin{equation}
R_{\rm II} = \int_{5.4~\mathrm{keV_{nr}}}^{225~\mathrm{keV_{nr}}}
\d T_\mathcal{N}^{\rm reco}\;
\frac{\d R}{\d T_\mathcal{N}^{\rm reco}}\bigg|_X\,.
\label{eq:RII}
\end{equation}

\section{Standard astrophysical neutrinos (solar, DSNB, and atmospheric)} 
As anticipated from the kinematic considerations discussed above, solar neutrinos cannot populate the signal region, while the DSNB contribution at the required energies is negligible. Atmospheric neutrinos therefore provide the only appreciable standard contribution to \(R_{\rm I}\), whereas all three components contribute to \(R_{\rm II}\). Within the SM, we find \(R_{\rm I}\simeq2\times10^{-6}\). Since standard astrophysical neutrinos interacting through SM CE\(\nu\)NS cannot account for LZ230616, we are motivated to explore whether new interactions could enhance these fluxes sufficiently to reproduce the LZ230616 event.

\begin{figure}[ht!]
\centering
\includegraphics[width=\linewidth]{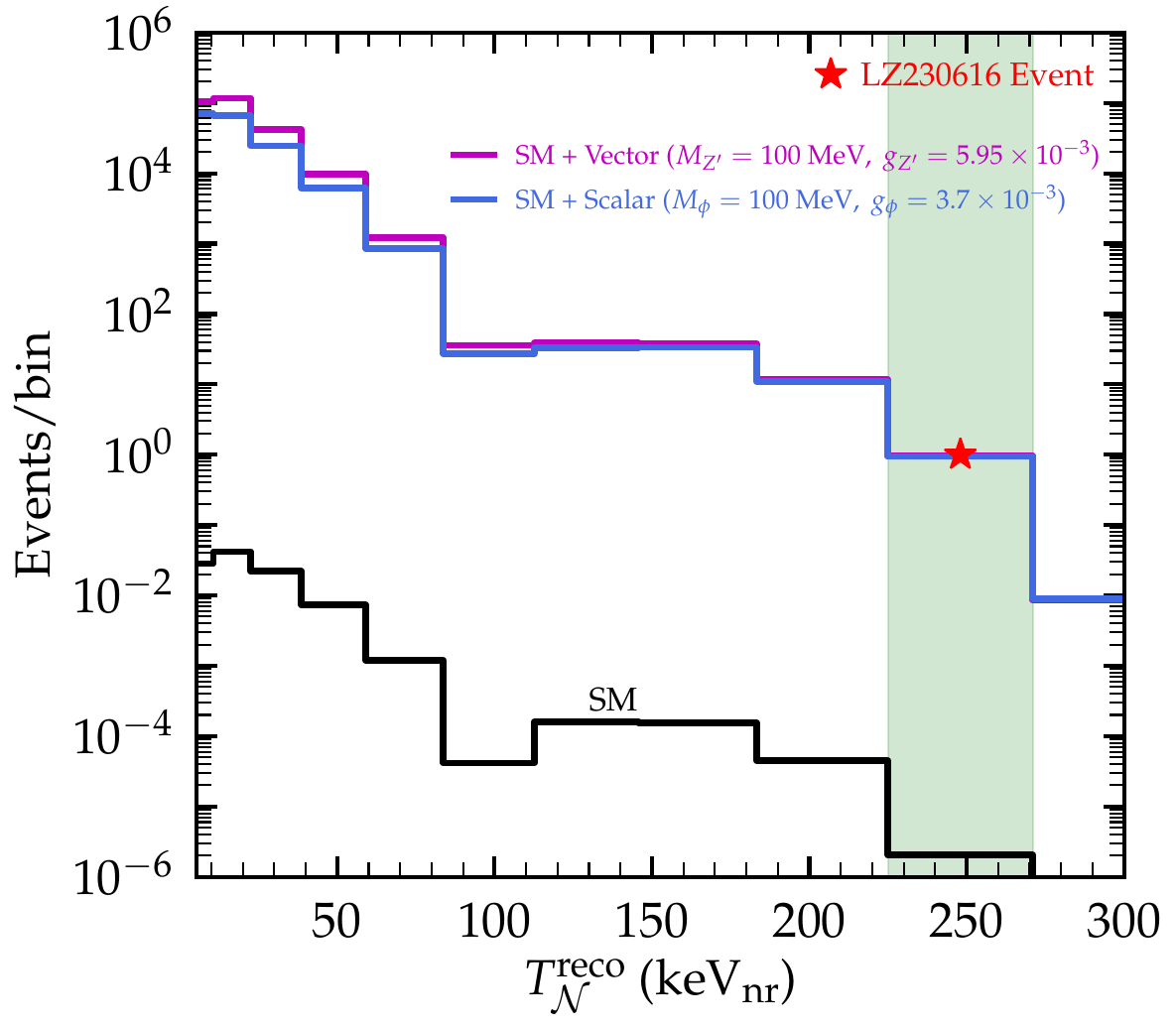}
\caption{\label{fig:BSM-events} Nuclear recoil event spectra induced by the conventional
astrophysical neutrino fluxes within the SM (black) and in the presence of
a novel vector (magenta) or scalar (blue) mediator, tuned so
that the signal region contains one event. The green shaded band indicates the signal
region and the red star marks LZ230616.}
\end{figure}

 Figure~\ref{fig:BSM-events} shows the recoil spectra
obtained in the presence of a vector or a scalar mediator, with the
couplings tuned so that the signal region contains one event. Due to the recoil dependence of the new physics cross sections, for both scenarios obtaining one event in the signal region requires a very large enhancement of the scattering rate. This enhancement is not localized around \(248~\mathrm{keV_{nr}}\) but extends throughout the recoil spectrum, producing \(\mathcal{O}(10^5)\) events in the lower-energy bins. Such a rate exceeds the SM prediction by about six orders of magnitude and is incompatible with the observed lower-energy spectrum, which is consistent with the expected background. Scanning over
$(M_{Z^\prime, \phi},~\textsl{g}_{Z^\prime, \phi})$ confirms that the $R_{\rm I} = 1$ contour
lies everywhere inside the region excluded by $R_{\rm II}$, for both
mediator types. Standard astrophysical neutrinos, whether scattering elastically
through the SM or through a new mediator, therefore do not provide a viable explanation.

\section{Neutrino flux from DM annihilation and decay}

Galactic DM can generate a monochromatic neutrino signal either through pair
annihilation or through decay. For DM annihilation into a neutrino pair with a thermally averaged annihilation cross
section $\langle \sigma v \rangle$, the combined neutrino and antineutrino flux reaching the Earth is~\cite{Yuksel:2007ac, Arguelles:2019ouk}
\begin{equation}
\frac{d\Phi_{\nu+\bar{\nu}}}{dE_\nu} =
\frac{\mathcal{J}}{4\pi}\,\frac{\langle \sigma v \rangle}{\kappa\, m_\chi^{2}}\,
\frac{1}{3}\,\frac{dN_\nu}{dE_\nu}\,,
\label{eq:flux-ann}
\end{equation}
where we set \(\kappa=2\), as appropriate for Majorana DM. 
For the case of DM decay with lifetime \(\tau_\chi\), the corresponding flux is~\cite{Palomares-Ruiz:2007egs, Arguelles:2022nbl}
\begin{equation}
\frac{d\Phi_{\nu+\bar{\nu}}}{dE_\nu} =
\frac{\mathcal{D}}{4\pi}\,\frac{1}{\tau_\chi m_\chi}\,
\frac{1}{3}\,\frac{dN_\nu}{dE_\nu}\,.
\label{eq:flux-dec}
\end{equation}
In both cases we have assumed flavor equipartition at Earth, so that the factor \(1/3\) converts the total neutrino yield into the yield per active flavor, an  approximation that is motivated by flavor mixing averaged over Galactic propagation distances.

We assume that DM annihilates (or decays) exclusively into a
neutrino pair, $\chi\chi \to \nu\bar{\nu}$ (or $\chi \to \nu\bar{\nu}$), yielding
the monochromatic neutrino spectrum
\begin{equation}
\frac{dN_\nu}{dE_\nu} =
\begin{cases}
\; 2\,\delta\!\left(1 - \dfrac{E_\nu}{m_\chi}\right)\dfrac{m_\chi}{E_\nu^{2}}
& \text{(annihilation)}\,, \\[3ex]
\; 2\,\delta\!\left(1 - \dfrac{2E_\nu}{m_\chi}\right)\dfrac{m_\chi}{2E_\nu^{2}}
& \text{(decay)}\,,
\end{cases}
\label{eq:spectrum}
\end{equation}
by noting that in both cases the spectrum is normalized to two
neutrinos per annihilation (decay), $\int dE_\nu\,(dN_\nu/dE_\nu) = 2$.

The astrophysical dependence is encoded in the line of sight integrals of the
dark matter density $\rho_\chi$, integrated over the solid angle $\Delta\Omega$
of the observed region,
\begin{align}
\mathcal{J} &= \int_{\Delta\Omega} d\Omega
\int_{\rm l.o.s.} ds\; \rho_\chi^{2}\big(r(s,\Omega)\big)\,,
\label{eq:Jfactor} \\[1ex]
\mathcal{D} &= \int_{\Delta\Omega} d\Omega
\int_{\rm l.o.s.} ds\; \rho_\chi\big(r(s,\Omega)\big)\,,
\label{eq:Dfactor}
\end{align}
with $r(s,\Omega) = \sqrt{s^{2} + R_\odot^{2} - 2\,s\,R_\odot \cos\Omega}$,
where $R_\odot$ is the distance of the Sun from the Galactic center. The Navarro-Frenk-White (NFW)
profile~\cite{Navarro:1995iw}
\begin{equation}
\rho_\chi(r) = \frac{\rho_s}
{\left(\dfrac{r}{r_s}\right)\left(1 + \dfrac{r}{r_s}\right)^{2}}
\label{eq:nfw}
\end{equation}
has been adopted which is normalized to the local dark matter density
$\rho_\odot = 0.4~ \mathrm{GeV\,cm^{-3}}$ at $R_\odot = 8.5~ \mathrm{kpc}$,
with scale radius $r_s = 20 ~ \mathrm{kpc}$. Integrating over the full sky,
yields
\begin{equation}
\begin{aligned}
    \mathcal{J} =& 1.58\times10^{23} \mathrm{GeV^{2}\,cm^{-5}\,sr}\, , \\
\mathcal{D} =& 2.79\times10^{23} \mathrm{GeV\,cm^{-2}\,sr}\,.
\end{aligned}
\label{eq:JD-values}
\end{equation}
 
\begin{figure*}[ht!]
\centering
\includegraphics[width=0.45\textwidth]{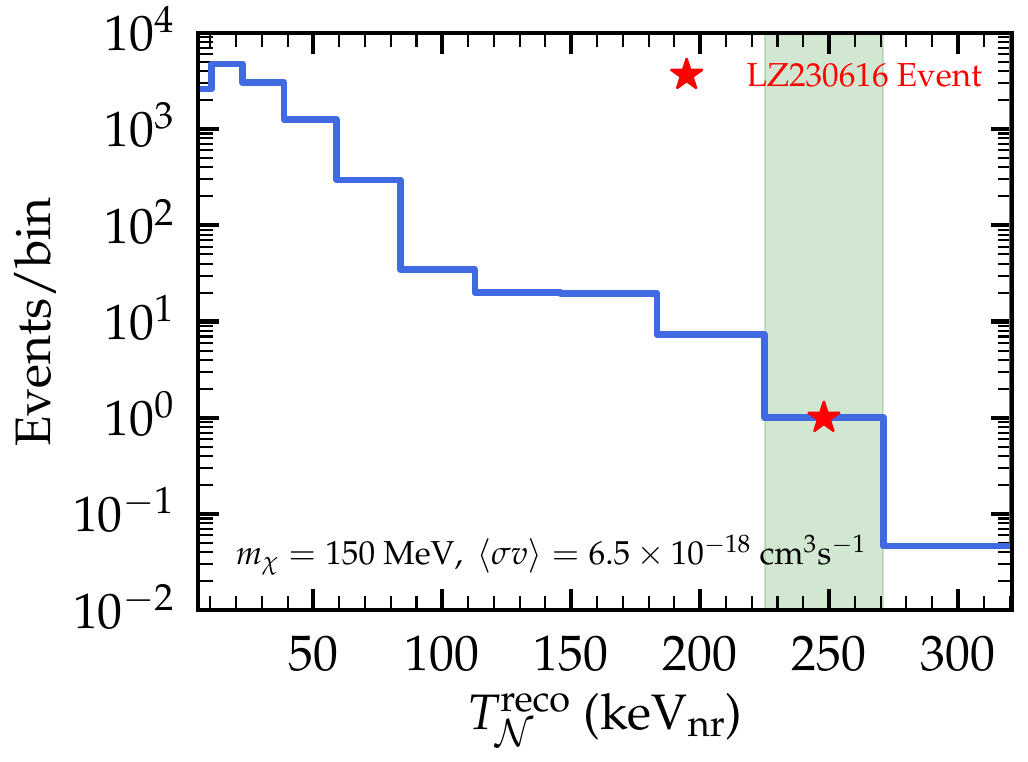}
\hspace{0.02\textwidth}
\includegraphics[width=0.46\textwidth]{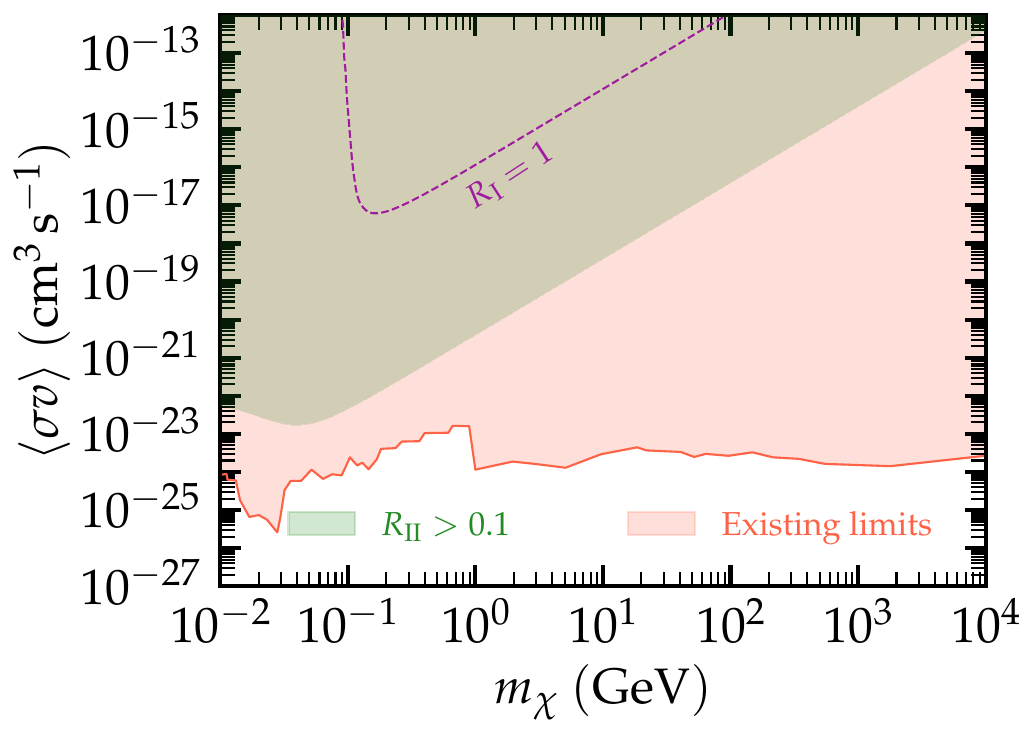}
\caption{\label{fig:DM-ann} \textit{Left:} SM recoil spectrum induced
by neutrinos from DM annihilation, for the benchmark
$m_\chi = 150~\mathrm{MeV}$ and $\langle \sigma v \rangle =
6.5\times10^{-18}~\mathrm{cm^3\,s^{-1}}$, chosen so the signal region contains one event. The red star marks LZ230616.
\textit{Right:} Exclusion regions in the $(m_\chi, \langle \sigma v \rangle)$
plane, illustrating the the $R_{\rm I} = 1$ contour (dashed magenta), and the excluded region by the $R_{\rm II} > 0.1$ condition (green
shaded area). Existing limits  are also superimposed~\cite{Arguelles:2019ouk}.}
\end{figure*}

Using the fluxes of Eqs.~(\ref{eq:flux-ann})-(\ref{eq:flux-dec})
and the event rate expression of Eq.~(\ref{eq:rate}), we examine whether either channel can
account for LZ230616. The left panel of Fig.~\ref{fig:DM-ann} shows the
SM recoil spectrum for annihilating DM, with $m_\chi$ and
$\langle \sigma v \rangle$ tuned so that the signal region contains one
event. The normalization required to reproduce one event in the signal region simultaneously predicts several hundred events at lower recoil energies, in clear conflict with the observed LZ spectrum. 

This tension is quantified in the right panel of Fig.~\ref{fig:DM-ann}.
Every point on the magenta dashed curve satisfies $R_{\rm I} = 1$, 
whereas the green region is where the low energy tail is overpopulated, $R_{\rm II} > 0.1$. The magenta curve lies
entirely inside the green region: there is no choice of
$(m_\chi, \langle \sigma v \rangle)$ capable of producing the LZ230616
event without simultaneously overproducing events at low recoil energies.
Moreover,  the annihilation cross sections required to reach
$R_{\rm I} = 1$ are of order
$\langle \sigma v \rangle \gtrsim 10^{-18}~\mathrm{cm^3\,s^{-1}}$, i.e.,
several orders of magnitude above existing bounds on DM annihilation into neutrinos. 

\begin{figure*}[ht!]
\centering
\includegraphics[width=0.45\textwidth]{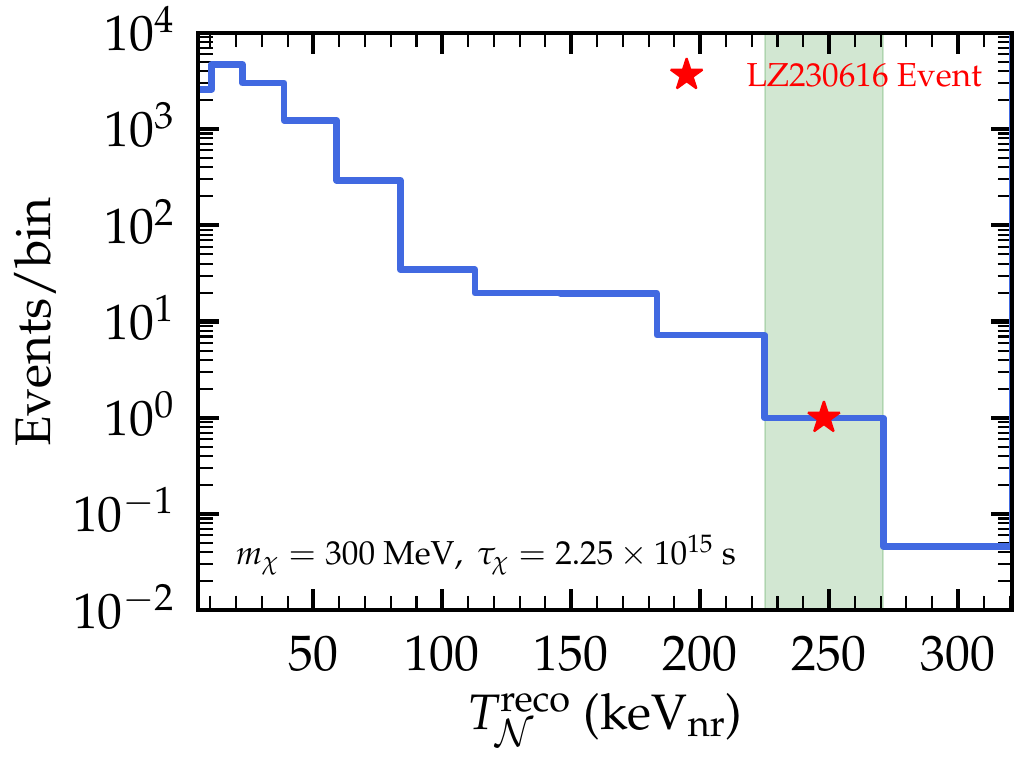}
\hspace{0.02\textwidth}
\includegraphics[width=0.47\textwidth]{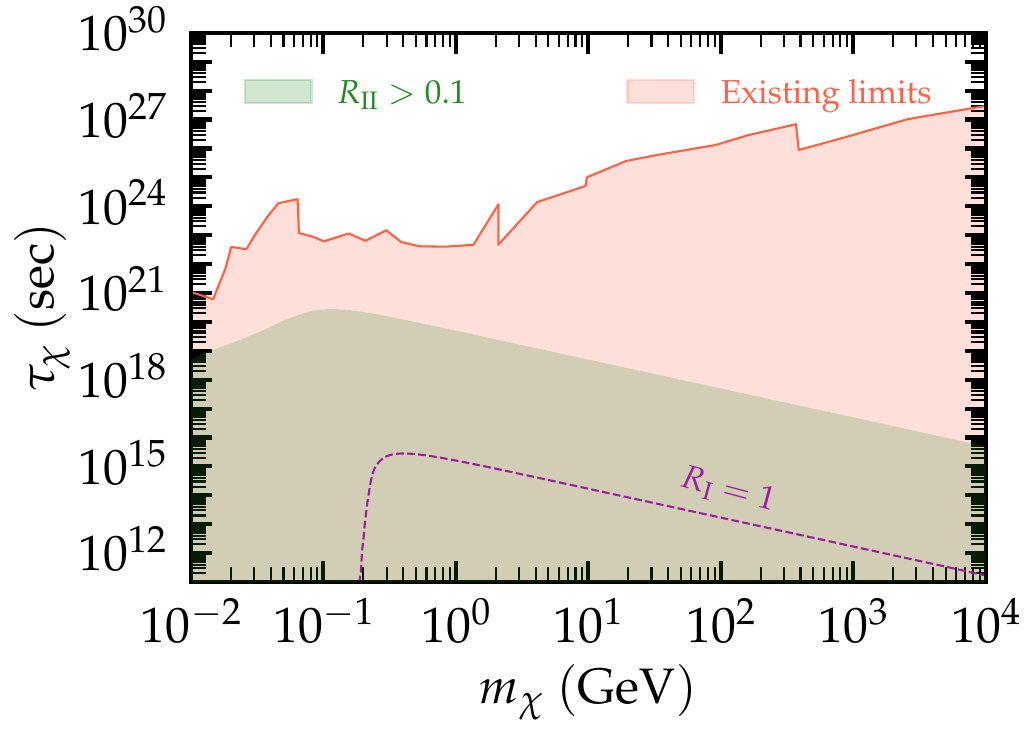}
\caption{\label{fig:DM-dec} Same as Fig.~\ref{fig:DM-ann}, but for
decaying DM. The left panel corresponds to the benchmark
$m_\chi = 300~\mathrm{MeV}$ and $\tau_\chi = 2.25\times10^{15}~\mathrm{s}$,
and the right panel shows the excluded region in the $(m_\chi, \tau_\chi)$ plane. In the right panel existing limits from different experiments are also shown~\cite{Arguelles:2022nbl}.}
\end{figure*}

The same conclusion applies to decaying DM, shown in Fig.~\ref{fig:DM-dec}.
The benchmark shown in the left panel yields the required one event in the signal region
but it also leads to an overabundance at low recoil energies. In
the right panel, the $R_{\rm I} = 1$ contour is found to  lie entirely within
the region excluded by the condition $R_\text{II}>0.1$. Likewise, the required lifetimes are
$\tau_\chi \lesssim 10^{15}~\mathrm{s}$, many orders of magnitude lower than
existing constraints already established from neutrino line searches. We therefore conclude that neutrinos from DM annihilation or decay, scattering elastically off target nuclei in the detector, cannot explain LZ230616.

\section{Neutrino flux from PBH evaporation} 
Another possibility is that primordial black holes (PBHs) constitute part of the dark sector and emit all particle species lighter than their temperature through Hawking radiation~\cite{Hawking:1974rv, Hawking:1975vcx}
\begin{equation}
T_{\rm PBH} = \frac{M_P^2}{M_{\rm PBH}}
\simeq 10~\mathrm{MeV}\left(\frac{10^{15}~\mathrm{g}}{M_{\rm PBH}}\right)\,,
\label{eq:TBH}
\end{equation}
with $M_P$ being the reduced Planck mass. The primary emission rate of neutrinos
with $g_\nu$ degrees of freedom follows a quasithermal
spectrum~\cite{Hawking:1974rv, Hawking:1975vcx}
\begin{equation}
\frac{d^2N_\nu}{dE_\nu \, dt}\bigg|_{\rm prim}
= \frac{g_\nu}{2\pi}\,
\frac{\Gamma_\nu(E_\nu, M_{\rm PBH})}{e^{E_\nu/T_{\rm PBH}} + 1}\,,
\label{eq:hawking}
\end{equation}
where $\Gamma_\nu$ is the greybody factor. In addition, decays and
hadronisation of the other primary products generate a secondary neutrino
component, which dominates at low energies. We evaluate the total rate
$d^2N_\nu/dE_\nu dt$ (primary plus secondary) with
\texttt{BlackHawk}~\cite{Arbey:2019mbc, Arbey:2021mbl}, assuming
nonrotating, uncharged (Schwarzschild-type) PBHs and a monochromatic mass function.
PBHs with $M_{\rm PBH} \gtrsim M_{\rm evap} \simeq
8\times10^{14}~\mathrm{g}$  survive until the present epoch and constitute a
fraction $f_{\rm PBH} \equiv \Omega_{\rm PBH}/\Omega_{\rm DM}$ of the dark
matter today. We restrict ourselves to this mass range, so that the signal
is controlled by the parameters $(M_{\rm PBH}, f_{\rm PBH})$.

The flux at Earth receives a galactic and an extragalactic contribution,
$d\Phi_\nu/dE_\nu = d\Phi^{\rm MW}_\nu/dE_\nu + d\Phi^{\rm EG}_\nu/dE_\nu$.
Unevaporated PBHs in the Milky Way halo trace the DM distribution, so the
galactic piece is again controlled by the $\mathcal{D}$-factor of
Eq.~(\ref{eq:Dfactor})~\cite{Bernal:2022swt},
\begin{equation}
\frac{d\Phi^{\rm MW}_\nu}{dE_\nu}
= \frac{\mathcal{D}}{4\pi}\,\frac{f_{\rm PBH}}{M_{\rm PBH}}\,
\frac{d^2N_\nu(E_\nu, t_0)}{dE_\nu \, dt}\,,
\label{eq:flux-pbh-mw}
\end{equation}
evaluated with the same NFW profile and normalization as above. The extragalactic component is obtained by integrating the emission over cosmic history while accounting for redshift~\cite{Bernal:2022swt}
\begin{equation}
\frac{d\Phi^{\rm EG}_\nu}{dE_\nu}
= \frac{f_{\rm PBH}\,\rho_{\rm DM}}{M_{\rm PBH}}
\int_0^{z_{\rm max}} \!\! dz \, \left|\frac{dt}{dz}\right|
\frac{d^2N_\nu\big(E_\nu(1+z), t\big)}{dE_\nu \, dt}\,,
\label{eq:flux-pbh-eg}
\end{equation}
where $\rho_{\rm DM} = \Omega_{\rm DM}\rho_c$ is the present DM energy
density and $|dt/dz| = \big[(1+z)H_0\sqrt{\Omega_\Lambda +
\Omega_m(1+z)^3}\big]^{-1}$.

Unlike the monochromatic lines of Eq.~(\ref{eq:spectrum}), the PBH neutrino spectrum is broad, with characteristic energies of order a few \(T_{\rm PBH}\) and an exponentially suppressed high-energy tail. Since \(T_{\rm PBH}\propto M_{\rm PBH}^{-1}\), lighter PBHs produce harder neutrino spectra, with a larger fraction of the emitted neutrinos satisfying \(E_\nu\gtrsim118~\mathrm{MeV}\). Finally, the primary Hawking spectrum is flavor universal. Neutrino oscillations can redistribute the flux among flavors but leave the total active neutrino flux unchanged~\cite{Bernal:2022swt}; we therefore work directly with Eqs.~(\ref{eq:flux-pbh-mw})--(\ref{eq:flux-pbh-eg}).

\begin{figure*}[ht!]
\centering
\includegraphics[width=0.45\textwidth]{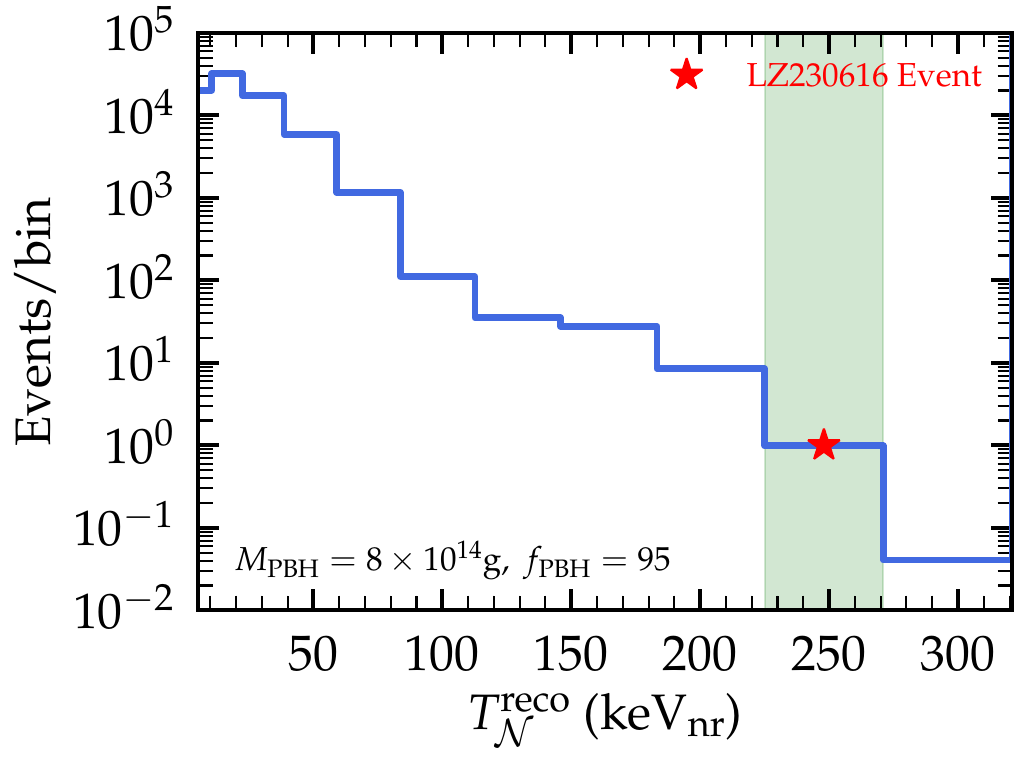}
\hspace{0.02\textwidth}
\includegraphics[width=0.47\textwidth]{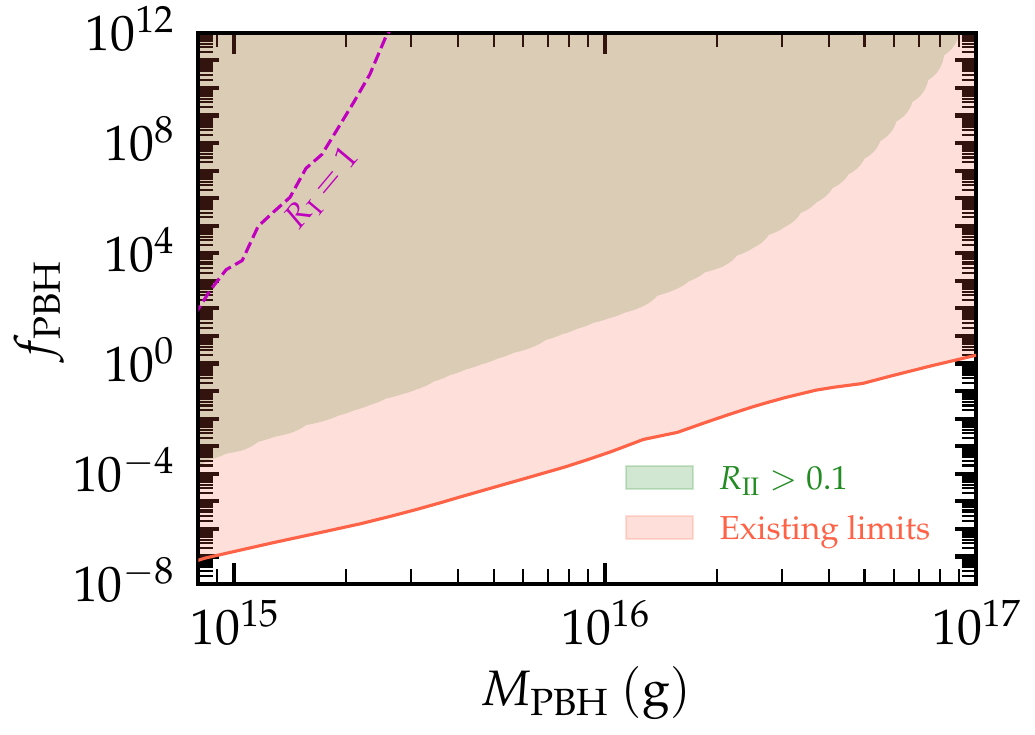}
\caption{\label{fig:PBH} \textit{Left:} SM recoil spectrum induced by
neutrinos from PBH evaporation, for the benchmark
$M_{\rm PBH} = 8\times10^{14}~\mathrm{g}$ and $f_{\rm PBH} = 95$, chosen so
that the highest bin accommodates exactly one event. The red star marks
LZ230616. \textit{Right:} Excluded region in the 
$(M_{\rm PBH}, f_{\rm PBH})$ parameter space, where the $R_{\rm I} = 1$ contour (dashed magenta) and $R_{\rm II} > 0.1$ region
(green shaded area) regions are depicted together with existing limits (red shaded area)~\cite{Carr:2020gox}.}
\end{figure*}

The left panel of
Fig.~\ref{fig:PBH} shows the recoil spectrum for the lightest PBH mass we
consider, for which we fix the $f_{\rm PBH}$ so that the signal region contains one event. 
Again, this normalization populates the lower recoil bins with $\mathcal{O}(10^4)$ events. The same kinematic argument applies, but the tension is strengthened by the shape of the PBH neutrino spectrum. Since flux decreases exponentially
above $E_\nu \sim \mathcal{O}(\mathrm{few}\,T_{\rm PBH})$, the neutrinos
energetic enough to deposit $\sim 248~\mathrm{keV_{nr}}$ sit far out in the
exponential tail, whereas the bulk of the emission produces recoils well
below the LZ230616 region. The resulting ratio $R_{\rm II}/R_{\rm I}$ grows
from $\sim 10^{4}$ at $M_{\rm PBH} \simeq 8\times10^{14}~\mathrm{g}$ to
$\sim 10^{23}$ for $M_{\rm PBH} \gtrsim 6\times10^{15}~\mathrm{g}$, as
colder PBHs emit an ever smaller fraction of their neutrinos at the
energies required.

As shown in the right panel of Fig.~\ref{fig:PBH}, the
$R_{\rm I} = 1$ curve lies entirely within the green region throughout the
mass range. This interpretation is also excluded independently by the required PBH abundance. Even at the lowest masses considered, obtaining $R_{\rm I}=1$ requires $f_{\rm PBH}\gtrsim95$, with the required fraction increasing steeply with $M_{\rm PBH}$. It therefore exceeds the total DM abundance by at least two orders of magnitude. The $R_{\rm I} = 1$ contour therefore never
approaches the physical region $f_{\rm PBH} \leq 1$, and it lies far above
the existing bounds on the PBH abundance as well. We thus conclude that
neutrinos from PBH evaporation, scattering elastically off target nuclei in
the detector, cannot account for LZ230616 either.

\section{Conclusions} 
The recent LZ observation of a single nuclear recoil at
$248 \pm 23~\mathrm{keV_{nr}}$, in a region where the expected background
is very low, raises the question of whether neutrinos rather than dark
matter could be responsible. We examined this
possibility for standard astrophysical fluxes
as well as for neutrinos of exotic origin such as those arising from DM annihilation and decay, and from the Hawking evaporation of PBHs.
For each source we required the predicted rate to reproduce one event in
the bin containing LZ230616, and tested its compatibility with the absence of an excess at lower
recoil energies.

None of the scenarios considered is viable: any neutrino flux capable of producing the \(\sim250~\mathrm{keV_{nr}}\) event inevitably overpopulates the more sensitive low-recoil region.  The resulting excess always exceeds the signal in the entire parameter space, for both DM annihilation and decay. The excess is even larger for PBHs, whose emission spectrum is exponentially suppressed above $\sim T_{\rm PBH}$. 
Moreover, the required parameter values are independently excluded: the DM annihilation cross sections lie above existing upper limits, the DM decay lifetimes fall below existing lower limits, and the required PBH abundance exceeds the total dark matter density of the Universe. Standard
astrophysical neutrinos likewise fail to account for the event through
either SM interactions or new scalar and vector mediators.
We therefore exclude elastic neutrino scattering from the sources considered here as an explanation of LZ230616, owing to the broad recoil spectrum inherent to elastic scattering kinematics.

\acknowledgments
D.K.P. acknowledges support from the European Union’s Horizon Europe research and innovation programme under the Marie Skłodowska‑Curie Actions grant agreement No.~101198541 (neutrinoSPHERE). D.K.P. benefits from the scientific environment of the Cluster of Excellence "Quantum Universe" funded by the Deutsche Forschungsgemeinschaft (DFG, German Research Foundation) under Germany’s Excellence Strategy (EXC 2121 “Quantum Universe”-390833306).

\bibliographystyle{utphys}
\bibliography{bibliography}

\end{document}